\documentclass[11pt]{article}
\usepackage{moriond,epsfig}

\def\be{\begin{equation}}
\def\ee{\end{equation}}
\def\bea{\begin{eqnarray}}
\def\eea{\end{eqnarray}}

\newcommand{\met} {\hbox{$E$\kern-0.55em\lower-0.15ex\hbox{/}}_T}

\begin{document}
\vspace*{4cm}
\title{TOP QUARK PAIR PRODUCTION CROSS SECTION AT THE TEVATRON}

\author{ GIORGIO CORTIANA }

\address{on behalf of CDF and $D\O$ collaborations \\
	 INFN \& University of Padova \\
          Department of Physics ``G. Galileo'' \\ 
	  Via Marzolo 8, 35131 Padova, Italy}

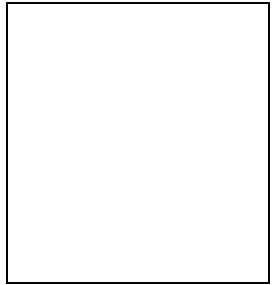
\maketitle\abstracts{
Top quark pair production cross section has been measured
at the Tevatron by CDF and $D\O$ collaborations using different
channels and methods, in order to test standard model predictions,
and to search for new physics hints affecting the $t\bar t$ 
production mechanism or decay. Measurements are carried out
with an integrated luminosity of 1.0 to 2.0 fb$^{-1}$, 
and are found to be consistent with standard model expectations.}

\section{Introduction}

\noindent
At the Tevatron $p\bar p$ collider top quarks are produced mainly in pairs
through quark-antiquark annihilation ($\sim 85\%$) and gluon-gluon fusion 
($\sim 15\%$) processes.
In the standard model (SM) the calculated cross section for pair production
is $6.7^{+0.7}_{-0.9}$ pb \cite{topthxsec} for a reference top mass of $175$ GeV/$c^2$,
and varies linearly with a slope of $-0.2$ pb/GeV/$c^2$ with the top quark
mass in the range $170$ GeV/$c^2 < m_t < 190$ GeV/$c^2$.
Accurate measurements of the $t\bar t$ production cross section serve as
important test of QCD calculation, and provide probes toward new 
physics signals involving non-SM production processes or $t\bar t$ decays.
Because the 
CKM element $V_{tb}$ is close
to unity and $m_t$ is large, the SM top quark decays  to a $W$ boson
and a $b$ quark almost $100\%$ of the time. The final state of top quark pair
production thus includes two $W$ bosons and two $b$-quark jets.
%
The $t\bar t$ experimental signatures are in general classified into three main 
categories:
the {\em di-lepton} category represents the case in which both $W$ bosons decay
leptonically; the {\em lepton+jets} signature arises when
one of the $W$ decays hadronically and the other into $l\nu_l$ (where $l=e,\mu,\tau$); 
finally, the {\em all-hadronic} channel corresponds to the case in which both 
$W$ bosons decay into quarks. 
%
In this proceeding a review of cross section measurements in various decays
channels and using different methods will be given, followed by a short 
description of some related searches for new physics hints.
%


\section{Top pair production cross section measurements}
\subsection{The dilepton channels}

\noindent
The $t\bar t$ dilepton channel accounts for $10.3\%$ of the total SM branching 
ratio. The experimental signature for these decays consist of two high-$p_T$, 
opposite sign leptons ($p_T\ge 15$ GeV/$c$), missing transverse energy 
($\met \ge 20\div 30$ GeV) and two or more high-$E_T$ jets from $b$-quark 
hadronization ($E_T \ge 15\div 20$ GeV).
The physics backgrounds to the $t\bar t$ signal are due mainly to $Z/\gamma^*$+jets and
diboson production ($WW/WZ/ZZ$+jets), and are estimated from Monte Carlo simulation.
Instrumental backgrounds are generated by lepton mis-identification ($i.e.$ from badly
reconstructed jets), artificial $\met$ from detector resolution effects or mis-measured
jets, and by mis-identified $b-$jets in analyses exploiting $b$-tagging techniques.
These background are estimated directly from data.

$D\O$ reports a combination of two analyses in the dilepton channel using 
$1.05$ fb$^{-1}$. In the standard dilepton analysis two well-identified and isolated
leptons are used to discriminate the $t\bar t$ signal from
backgrounds. On the other hand, the so-called lepton+track analysis relaxes 
the identification requirements on the second lepton to an isolated track and 
restores the signal to background ratio by requiring $b$-jets identification. 
In this way lepton+track selected events increase by $\sim 30\%$ the total 
$t\bar t$ acceptance, in particular with respect to $W\to\tau\nu_\tau$ 
decays. The combined dilepton and lepton+track samples provide a cross section
measurement of $\sigma_{t\bar t} = 6.2 \pm 0.9$(stat.) $\pm 0.7$(syst.) $\pm 0.5$ (lumi.) pb 
for $m_t = 175$ GeV/$c^2$, with a total relative uncertainty of
$19\%$ \cite{d0dil}.

CDF reports three different measurements in the dilepton channel:
$\sigma_{t\bar t} = 6.2 \pm 1.0$(stat.) $\pm 0.7$(syst.) $\pm 0.4$(lumi.) pb, using
a data sample selected by requiring two well identified $e$ or $\mu$; 
$\sigma_{t\bar t} = 8.3 \pm 1.3$(stat.) $\pm 0.7$(syst.) $\pm 0.5$(lumi.) pb, and
$\sigma_{t\bar t} = 10.1 \pm 1.8$(stat.) $\pm 1.1$(syst.) $\pm 0.6$(lumi.) pb, in 
the lepton+track and and lepton+track plus $b$-tag channels respectively \cite{cdfdil}.

In addition to the traditional analyses involving $e/\mu$ lepton pairs outline above, 
$D\O$ reports measurements in the $e/\mu +\tau$ channel, which are of particular 
interest to look for effect beyond the SM, like those involving 
$t\to H^+b \to \tau \nu_\tau b$ decays.  
Events are selected requiring one well identified $e$ or $\mu$, and a $\tau$ candidate 
satisfying a Neural Network selection optimized for $\tau$-hadronic decays. In 
addition, at least one jets is required to be identified as originating from $b$-quark.
The $t\bar t$ cross section is measured assuming SM branching ratios, including
extra signal acceptance from $t\bar t$ events in the dilepton or lepton+jets channel in
which one electron or a jets mimics the $\tau$-signature. The cross section is 
measured to be 
$\sigma_{t\bar t} = 8.3^{+2.0}_{-1.8}$(stat.) $^{+1.4}_{-1.2}$(syst.) $\pm 0.5$(lumi.) pb.
On the other hand, assuming the SM cross section, the values of the 
exclusive $\sigma_{t\bar t}\times BR(t\bar t\to e+\tau+2\nu+2b)$ and
$\sigma_{t\bar t}\times BR(t\bar t\to \mu+\tau+2\nu+2b$) are found to be
$0.19 ^{+0.12}_{-0.10}$(stat) $\pm 0.07$(syst.) pb, and
$0.18 ^{+0.13}_{-0.11}$(stat) $\pm 0.09$(syst.) pb respectively \cite{d0taulep}. 
These measurements are in good agreement with the SM expectation of 0.13 pb 
and 0.12 pb for the $e+\tau$ and $\mu+\tau$ channels respectively. 

\subsection{The Lepton+jets channel}

\noindent
Lepton+jets $t\bar t$ decays account for up to $43.5\%$ of the total branching ratio when
all lepton flavors are considered, and it
is widely used for top-quark properties measurements.

In general, except for the results outlined in \cite{cdfmetjets},
the basic event selection requires one well-identified high-$p_T$ $e$ or $\mu$ 
($p_T \ge 20$ GeV/$c$), missing transverse energy ($\met \ge 20\div 30$ GeV) and 
at least three high-$E_T$ jets ($E_T\ge 15 \div 20$ GeV), two of which can 
be required to satisfy $b$-tagging requirements.
The dominant physics background originates from $W$+jets production, with small
contributions from $Z\to\tau\tau$, single top and diboson production processes. 
Instrumental background is mainly due to fake isolated 
leptons in multijet events. Background contributions are calculated from 
Monte Carlo simulation and data.

To separate the $t\bar t$ signal from background contributions two approaches
are followed. The first, in addition to the core selection outlined above, adopts
$b$-jets identification algorithms to increase the sample purity. The second
one, since no single kinematical variables provides sufficient discrimination
power, makes use of a complex topological discriminant exploiting multiple kinematical
properties of the events, $i.e.$ the $\Delta R$ between leptons and jets, or between 
jets, the total event transverse energy ($H_T$), aplanarity, and sphericity.
The latter approach is of particular interest since it does not rely on
$b$-tagging techniques, and is free from the $BR(t\to Wb)\sim 1$ assumption.

$D\O$ reports results from a combination of these two techniques, 
using $0.9 fb^{-1}$, providing a $t\bar t$ cross section determination with a 
relative total uncertainty of $\sim 10\%$, at the same level of the theoretical
uncertainty
($\sigma_{t\bar t} = 7.42\pm 0.53$(stat.)$\pm 0.46$ (syst.) $\pm 0.45$(lumi.) pb 
\cite{d0ljcomb}).
Additionally a simultaneous measurement of the $t\bar t$ production cross
section and of the ratio $R=\frac{BR(t\to Wb)}{BR(t\to Wq)}$ provides the best
single measurement of the $\sigma_{t\bar t}$ to date: $8.18 ^{+0.90}_{-0.84}$
(stat.+syst.)$\pm 0.5$(lumi.) pb\cite{d0ljRs}. 

CDF reports different measurements of the $t\bar t$ cross section in the lepton+jets
channel as well. The analysis requiring at least one identified $b$-jet in addition
to the standard selection, measures 
$\sigma_{t\bar t} = 8.2\pm 0.5$(stat.) $\pm 0.8$(syst.) $\pm 0.5$(lumi.) pb
\cite{cdfljbtag} 
($\sim 13\%$ relative uncertainty). Other measurements using a topological discriminant
or a combined secondary vertex and Neural Network tagger are reported in 
\cite{cdfljother}, for a data sample of approximately $0.7$ fb$^{-1}$. 
The main sources of systematic uncertainty, as in the case of the dilepton channel,
are due to lepton identification, jet energy scale, background and Monte Carlo normalization,
and $b$-tagging efficiency.

\subsection{The All-hadronic channel}

The all-hadronic channel has the highest branching ratio ($46.2\%$), 
but suffers from the largest background contribution. 
All-hadronic $t\bar t$ events are characterized by the basic
signature of 6 or more high-$E_T$ jets, two of which originate from $b$-quark. 
The physics background is due to QCD multijet production, which can be reduced
by applying specific topological and kinematical Neural Network based selection in combination
with $b$-tagging requirements. 

CDF and $D\O$ report measurements\cite{cdfd0allhad} 
in this channel using $1.05$ fb$^{-1}$ and $0.4$ fb$^{-1}$ to be  
$\sigma_{t\bar t} = 8.3\pm 1.0$(stat.) $^{+2.0}_{-1.5}$(syst.) $\pm 0.5$(lumi.) pb,
and
$\sigma_{t\bar t} = 4.5^{+2.0}_{-1.9}$(stat.) $^{+1.4}_{-1.1}$(syst.) $\pm 0.3$(lumi.) pb,
respectively. The main systematics for these measurements is related to uncertainties
in the jet energy scale.

\section{Top quark production cross section properties, and probes toward new physics}


\subsection{Lepton+jets/dilepton cross section ratio}

The ratio between $t\bar t$ cross section measurements in the lepton+jets and 
dilepton channel is sensitive to non-SM top decays: top quark pair
decays involving a charged Higgs boson could enhance the observed lepton+jets
cross section ratio, while lowering the dilepton one. $D\O$ report 
a measurement of  $R=\frac{\sigma_{t\bar t}^{l+jets}}{\sigma_{t\bar t}^{dil}} = 1.21 ^{+0.27}_{-0.26}$,
in good agreement with the SM expectation of unity. 
Moreover, assuming a $m_{H^+}\sim m_W$, and $BR(H^\pm\to cs\sim 1$), a 95\% C.L. upper
limit is set at 0.35 for the $BR(t\to H^+b)$ \cite{d0ratio}.

\subsection{Top pair production properties and resonance searches}

Top quark pairs at the Tevatron center-of-mass energy ($\sqrt{s}= 1.96 TeV$) are
produced mainly through quark-antiquark annihilation and gluon-gluon 
fusion processes. The relative fraction of gluon-gluon production processes, $F_{GG}$,
is measured by CDF using $0.95$ fb$^{-1}$. The result is obtained by combining 
two methods using low-$p_T$ track multiplicity and kinematical variables to 
discriminate between the two production mechanisms. $F_{GG}$ is measured to
be $0.07^{+0.15}_{-0.07}$ (stat+syst) and is constrained to be less than
0.38 at $95\%$ C.L. \cite{cdfFgg}, in good agreement with the SM expected 
value of 0.15.

The $t\bar t$ invariant mass spectra obtained in the lepton+jets channel
is used by both collaborations to check for shape distortion effects
or for the presence of bump yielded by physics processes beyond the SM.
The technicolor $Z^\prime$ has been excluded at 95\% C.L. for masses 
below  760 GeV/$c^2$ (by $D\O$ using $2.1$ fb$^{-1}$) and 720 GeV/$c^2$
(by CDF using $0.95$ fb$^{-1}$) \cite{d0cdfbump}.

Additionally, distortion of the $t\bar t$ invariant mass shapes due to the 
interference between quark-quark annihilation processes mediated by standard
massless gluons and hypothetical massive ones has been checked by CDF using $1.9$ fb$^{-1}$. 
The overall agreement with respect to the SM, in terms of massive gluon coupling strength, 
is within 1.7 $\sigma$ for 
massive gluon masses, $M_G$, from 400 to 800 GeV/c$^2$, and widths in the 
range $[0.05; 0.5]M_G$\cite{cdfmg}.

Finally, the differential cross section $d\sigma_{t\bar t}/dM_{t\bar t}$, 
corrected for detector resolution effects using regularized unfolding techniques,
is measured using $1.9$ fb$^{-1}$, and is found to be consistent with SM expectation
with a probability of 45\% \cite{cdfmg}. 

\section{Conclusions}

The $t\bar t$ production cross section has been measured by CDF and $D\O$ 
collaborations in different channels and using different methods. 
Possible hints of physics beyond the SM affecting $t\bar t$ production or 
decays are investigated using the cross section information, searching for
shape distortions or bumps in the $t\bar t$ invariant mass spectra in the
lepton+jets channel, or measuring cross section ratio in different channels.
In all cases, current experimental measurements are in agreement with SM expectations.
Many updates are foreseen for the Summer 2008.

\section*{Acknowledgments}

I would like to thank CDF and $D\O$ collaborators, and top-group conveners
for their
suggestions and help. I would also like to thank Moriond QCD 2008 
organizers for the very pleasant, and at the same time interesting and 
professional conference atmosphere.

\section*{References}

\end{document}